\documentclass[reprint, 
aps,
amsmath,amssymb,floatfix,showkeys]{revtex4-2}

\usepackage{graphicx}
\usepackage{dcolumn}
\usepackage{bm}

\usepackage{cmap}					
\usepackage{mathtext} 				
\usepackage[T2A]{fontenc}			
\usepackage[utf8]{inputenc}			
\usepackage[main=english]{babel}	

\DeclareMathOperator*{\argmin}{arg\,min}
\renewcommand{\epsilon}{\ensuremath{\varepsilon}}
\renewcommand{\phi}{\ensuremath{\varphi}}
\renewcommand{\kappa}{\ensuremath{\varkappa}}

\renewcommand{\vec}[1]{\ensuremath\bm{#1}}
\renewcommand{\eqref}[1]{Eq.~(\ref{#1})}
\newcommand{\fref}[1]{Fig.~\ref{#1}}

\usepackage[bookmarks=false,breaklinks=true]{hyperref}
\usepackage[usenames,dvipsnames,svgnames,table,rgb]{xcolor}
\hypersetup{				
	unicode=true,           
	pdftitle={},   
	pdfauthor={Yuriy Kanygin},      
	pdfsubject={VF-DFT},      
	pdfcreator={Kanygin}, 
	pdfproducer={Kanygin}, 
	pdfkeywords={DFT} {PSO} {GA}, 
	colorlinks=true,       	
	linkcolor=red,          
	citecolor=black,        
	filecolor=magenta,
	urlcolor=black 
}

\usepackage{booktabs}
\usepackage{multirow}
\newcommand{\frc}[2]{\raisebox{2pt}{$#1$}\big/\raisebox{-3pt}{$#2$}}
\usepackage{float}

\usepackage{pstool}
 
\usepackage{adjustbox}

\usepackage{tikz}
\usetikzlibrary{shapes.geometric, arrows}

\tikzstyle{cdft} = [rectangle, rounded corners, minimum width=1cm, minimum height=1cm,text centered,text width=2.6cm, draw=black]

\tikzstyle{vfdft} = [rectangle, rounded corners, minimum width=1cm, minimum height=1cm,text centered,text width=2.6cm, draw=red]

\tikzstyle{empty} = [minimum width=1cm, minimum height=1cm,text centered,text width=1cm]

\tikzstyle{arrow} = [thick,->,>=stealth]

\usepackage[justification=centerlast]{caption}
\usepackage{subcaption}
\usepackage[section]{placeins}

\begin{document}

\preprint{}

\title{Variation Free Approach for Molecular Density Functional Theory:\\ Data-driven Stochastic Optimization}

\author{Yuriy Kanygin}
\email{yuriy.kanygin@phystech.edu}
\affiliation{%
Moscow Institute of Physics and Technology,\\
Center for Engineering and Technology of MIPT 
}
 
\author{Irina Nesterova}%
\email{irina.nesterova@phystech.edu}
\affiliation{%
Moscow Institute of Physics and Technology,\\
Center for Engineering and Technology of MIPT 
}

\author{Pavel Lomovitskiy}%
\email{pavel.lomovitskiy@phystech.edu}
\affiliation{%
Moscow Institute of Physics and Technology,\\
Center for Engineering and Technology of MIPT
}

\author{Aleksey Khlyupin}%
\email{khlyupin@phystech.edu}
\affiliation{%
Moscow Institute of Physics and Technology,\\
Center for Engineering and Technology of MIPT
}

\date{\today}
\begin{abstract}

Density functional theory (DFT) is an efficient instrument for describing a wide range of nanoscale phenomena: wetting transition, capillary condensation, adsorption, etc. In this paper, we suggest a method for obtaining the equilibrium molecular fluid density in a nanopore using DFT without calculating the free energy variation --- Variation Free Density Functional Theory (VF-DFT). This technique can be used to explore confined fluids with a complex type of interactions, additional constraints and to speed up calculations, which might be crucial in an inverse problems. The fluid density in VF-DFT approach is represented as a decomposition over a limited set of basis functions. We applied Principal Component Analysis (PCA) to extract the basic patterns from the density function and take them into account in the construction of a set of basis functions. The decomposition coefficients of the fluid density by the basis were sought by stochastic optimization algorithms: genetic algorithm (GA), particle swarm optimization (PSO) to minimize the free energy of the system. In this work, two different fluids were studied: nitrogen at a temperature of 77.4~K and argon 87.3~K, at a pore of 3.6~nm, and the performance of optimization algorithms was compared. We also introduce the Hybrid Density Functional Theory (H-DFT) approach based on stochastic optimization methods and the classical Picard iteration method to find the equilibrium fluid density starting from the physically appropriate solution. The combination of Picard iteration and stochastic algorithms helps to significantly speed up the calculations of equilibrium density in the system without losing the quality of the solution, especially in cases with the high relative pressure and expressed layering structure.

\end{abstract}
\keywords{Density Functional Theory, Stochastic Optimization, Principle Component Analysis, Genetic Algorithm, Particle Swarm Optimization}
\maketitle
\newpage

\section{Introduction}
\begin{figure*}[bh!t]
\centering
\resizebox{\textwidth}{!}{
\begin{tikzpicture}[node distance=3.5cm]
\node (start) [cdft,text width=2.85cm] {Fluid Properties\\ $F\left[\rho\left(z\right)\right]$};
\node (varcalc) [cdft,right of=start,xshift=1.5cm,yshift=1.3cm,align=left] {Variation Calculation};
\node [empty,right of=varcalc,xshift=-2.6cm] {$\dfrac{\delta F\left[\rho\right]}{\delta\rho\left(z\right)}$};

\node (basisfunc) [vfdft,right of=start,yshift=-1.3cm] {
\includegraphics[scale=0.8]{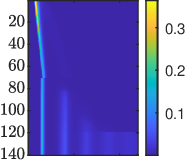}
Dataset $\bm{X}$};
\node (PCA) [vfdft,right of=basisfunc,text width=2.8cm] {
\includegraphics[scale=0.8]{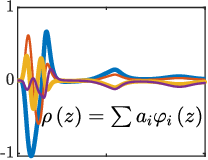}
Pattern extraction};
\node (stochopt) [vfdft,right of=PCA,text width=2.85cm,xshift=0.2cm] {Stochastic\\ Optimization\\ $\rho^*_{so}=\argmin\limits_{\widetilde{a}_i}\Omega\left(\widetilde{a}_i\right)$};

\node (picard) [cdft,right of=varcalc,xshift=3cm] {Picard Iteration};

\node (dens) [cdft,right of=stochopt,xshift=1cm,yshift=1.3cm] {Equilibrium Density $\rho^{*}\left(z\right)$};

\draw [arrow] (start) |- node[anchor=south west] {Classical DFT} (varcalc);
\draw [arrow,red] (start) |- node[anchor=north west, red] {VF-DFT} (basisfunc);
\draw [arrow,red] (basisfunc) -- (PCA);
\draw [arrow,red] (PCA) -- (stochopt);
\draw [arrow] (varcalc) -- (picard);
\draw [arrow] (picard) -- +(3.7,0) -- (dens);
\draw [arrow,red] (stochopt) -- +(4.5,0) -- (dens);
\draw [arrow,blue] (stochopt) -- +(0,2.1);
\draw [arrow,blue] (picard)+(0.6,-0.5) |- node[anchor=south east, blue] {H-DFT} (dens);
\end{tikzpicture}
}
\caption{Workflow for the classical DFT with the Picard iterations (black line), the Variation Free DFT method (red line), and the hybrid DFT (red and blue lines)}
\label{fig:TOC}
\end{figure*}
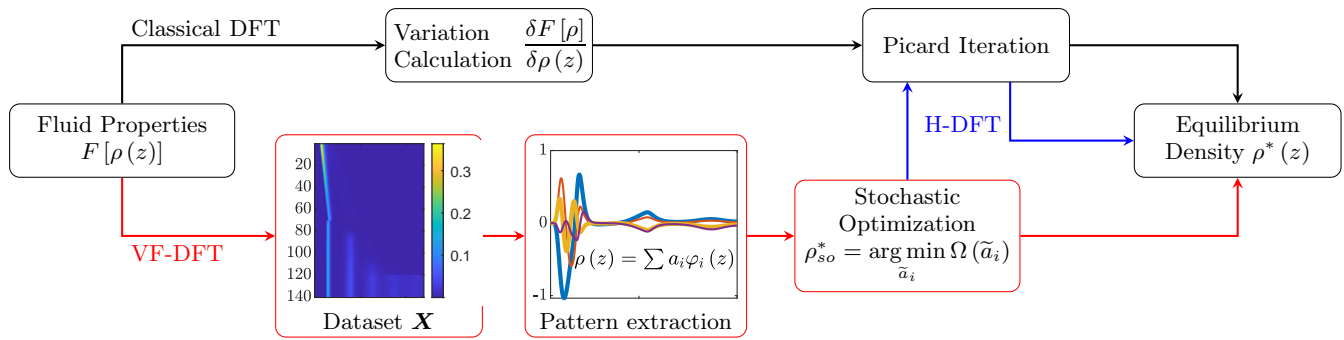 


Studying fluids in confinement at scales from micro- to nanometers might be important for many technological industries such as biomolecular engineering, fabrication of novel materials and the oil and gas industry \cite{RezaeiGomari2019NewReservoirs,Wang2018AtomisticCapillary,Neimark1998PoreAdsorption,Yu2004Density-functionalSolutions,muller2003wetting, bi2019molecular, Wu2006DensityMaterials}. For instance, for the petroleum industry in unconventional reservoirs, most of the pore space represent structures with a width of several nanometers \cite{zhang2017dft,RezaeiGomari2019NewReservoirs, bi2019molecular}. On such a scale, classical approaches have limitations, and it becomes necessary to take into account intermolecular forces more accurately to describe such physical phenomena as adsorbtion \cite{Neimark1998PoreAdsorption, Ravikovitch2001DensityNanopores}, capillary condensation \cite{Wu2006DensityMaterials}, wetting transition \cite{Berim2008NanodropConsiderations, Wu2006DensityMaterials}, heterogeneity of the reservoir walls \cite{Aslyamov2017DensitySurfaces,aslyamov2019random,aslyamov2019theoretical,Neimark2009QuenchedCarbons, Jagiello20132D-NLDFTCorrugation, Yang2011SolvationSurfaces} and so forth. An accurate description of these phenomena and the physical properties of materials are extremely important for industry.

Density functional Theory (DFT) is one of the most common tools for describing equilibrium thermodynamic and structural properties of fluids under different external potential \cite{evans1979nature,Wu2006DensityMaterials, Roth2010FundamentalReview,Neimark1998PoreAdsorption,Ravikovitch2000UnifiedIsotherms,Ravikovitch2001DensityNanopores,balbuena1993theoretical}. DFT provides a compromise between classical semi-empirical methods and molecular modeling. On the one hand, it is capable of taking into account the microscopic structure of a macroscopic system at relatively low computational costs. On the other hand, DFT is a more rigorous theory than classical phenomenological approaches.

According to the mathematical theorem on which the DFT is based, the variation of the free energy with respect to density must be zero as a necessary condition for the minimum  \cite{Wu2006DensityMaterials,Oxtoby2002DensityMaterials}, and thus, further calculations of the equilibrium density are based on this statement.

The Helmholtz free energy defines the physics of the system, and it has a complex structure for the real fluid. So the variation calculating might be a difficult issue for such systems. Therefore, it is necessary to construct various approximations that describe certain physical phenomena with precise accuracy. Recently developed approaches allowed taking into account the heterogeneity of the surface \cite{Aslyamov2017DensitySurfaces, Neimark2009QuenchedCarbons, Jagiello20132D-NLDFTCorrugation}, adsorbtion-induced deformation in non-convex materials \cite{ludescher2021adsorption}, or combine DFT with elasticity theories for studying graphene nanobubbles \cite{aslyamov2020model}. There are certain subtleties in the last application, fluid in such a system is under additional constrain, and it is crucial to know both the value of system energy and equilibrium density at the same time. The work \cite{Aslyamov2017DensitySurfaces} demonstrated that the account interaction between fluid molecules and the heterogeneous surface might significantly change the system's free energy and its variation. The most significant change undergone the part with the variation of attraction interaction. Depending on what material is being investigated, the variation of free energy has to be calculated in different coordinate systems \cite{Roth2010FundamentalReview,xi2020efficient, ludescher2021adsorption}, which means variations should be recalculated in appropriate coordinates. Moreover, Statistical
Associating Fluid Theory (SAFT) is usually used for describing polymers or other complex fluids, which considers dispersion, association, and chain contributions in Helmholtz energy \cite{xi2020efficient}. SAFT is an advanced technique for describing real fluids, but it has many modifications for which variations should be calculated.

Many applications require a large amount of direct computation, so it is important to perform these computations quickly. Nanoporous material such as zeolites, metal–organic frameworks, and covalent organic frameworks are promising as adsorbents with ultrahigh capacity and separation selectivity \cite{zhou2020gpu}. Porous polymers can be used as gas storage and separation materials as encapsulation agents for controlled release of drugs, as catalysts and so forth \cite{wu2012design}. DFT might be applied for the prediction of gas adsorbtion in nanomaterials and consequently for nanoporous material design. Moreover, DFT is usually applied for solving such inverse problem as pore size distribution reconstruction \cite{Ravikovitch2000UnifiedIsotherms}. For such inverse problems, it is important to make a large number of direct calculations. And our method might be applicable to speed it up. 

In this paper, we propose a new approach for calculating the equilibrium density without using the Helmholtz free energy variation. The algorithm of the VF-DFT is shown schematically in \fref{fig:TOC} compared with the classical DFT. At the initial stage, once for all subsequent tasks, a dataset of functions is calculated. The dataset contains the typical information about fluid density behavior in a nanopore. We used nitrogen densities (which were calculated using the classical approach with the Piсard iteration method) in carbon nanopore with width 3.6~nm under different relative pressures (bulk pressure relate to saturation pressure). Also, we added to the dataset "artificial densities" which were calculated with parameters similar to nitrogen, but with different molecular radii. Adding these "artificial densities" allowed us to add a variety to the dataset and applied our approaches to other fluids. As a result, we got a dataset of 140 different functions.

Since the confined fluid density has typical structural patterns, these patterns can be efficiently extracted from the densities dataset using principal component analysis (PCA). And, after that, those patterns might be used for calculating the set of basis functions. In all furthers calculations the desired density was represented as an expansion into the basis functions. Thus, finding an equilibrium density was reduced to the problem of optimizing the expansion coefficients. Based on the energy criterion  \cite{Sun2017AProblems,elizarev2020objective,mukhin2020application} for PCA, it was possible to decrease the number of basis functions and coefficients from 140 to 10. The resulting 10 vectors contain 95 \% information about the dataset, which means that we have approximately the same quality of a solution as with 140 coefficients.

In the Variation Free algorithm, the search for the density expansion coefficients is carried out using stochastic optimization methods, in contrast to the classical variational approaches, which calculate density based on the Picard iteration or Newton's method \cite{Roth2010FundamentalReview,Wu2017VariationalModeling,Edelmann2016ATheory,Sears2003AFluids, mairhofer2017numerical,zhou2020gpu}. Among the wide range of stochastic optimization algorithms, we compared two the most common and popular: genetic algorithm (GA) and particle swarm optimization (PSO). Stochastic optimization algorithms have the advantage that they are less affected by local minima and are well applicable for finding global ones. In particular, the algorithm produces a set of expansion coefficients for the density, which corresponds to the equilibrium state of the system and the minimum free energy. 

The hybrid algorithm H-DFT \fref{fig:TOC}
combines the advantages of both VF-DFT and classical DFT with Picard iterations. The quality of solutions for VF-DFT is determining by the type and number of basis functions, which were constructed once for all calculations. Because of this, VF-DFT solutions might be different from the classical DFT approach, but the calculation time for VF-DFT is significantly less. To clarify the solution of VF-DFT we were applying the classical DFT approach with the initial guess from the solution of VF-DFT. This combination still works faster than classic DFT almost for all our cases and the quality of solutions the same as the classical approach solution. However, H-DFT requires free energy variation calculations. Combining these methods reduces the time needed to find the equilibrium density without losing the quality of solutions because the rate of convergence of Picard iterations depends on the choice of the initial approximation. The closer it is to the solution, the faster the method converges. Thus, for H-DFT we need to use the full algorithm of VF-DFT and then set the initial guess for Picard iteration to clarify the solution.

The developed approach was applied to find the equilibrium fluid density in a pore of 3.6~nm. We considered: nitrogen and argon as fluids, carbon pore walls and planar geometry. The Steele potential 10-4-3 \cite{steele1974interaction} was taken as the solid-fluid interaction potential. To describe the fluid-fluid interaction we used FMT approach for hard-sphere repulsion, developed by Rosenfeld \cite{Rosenfeld1989Free-EnergyFreezing, Roth2010FundamentalReview} and WCA \cite{Weeks1971RoleLiquids} for the attraction interaction.

As will be shown in section \ref{sec:results}, pleasing results are obtained for nitrogen using VF-DFT and H-DFT in a significantly shorter time. In the best case, it turned out that the VF-DFT produced a solution 36 times faster than the classical DFT with Picard iterations. Even though the basis was built based on the fluid with nitrogen parameters, it is possible to calculate the equilibrium density of the fluid, information on which was not in the basis. VF-DFT coped well with the task of calculating the equilibrium density of argon.  

The article is organized as follow, in the next section we will provide a brief reference on the density functional theory and the Picard iteration method. Section \ref{sec:VF-DFT} will describe a method for basic patterns extraction from data on the density behavior confined system, optimization methods used in this work, and consider the advantages of VF-DFT and H-DFT approaches. Then, in section \ref{sec:results}, the results will be considered for nitrogen and argon equilibrium density obtained with the VF-DFT, H-DFT, and classical approach DFT with iterations Picard.

\section{Density Functional Theory}
The density functional theory is based on the fact that the free energy of the system is minimal in equilibrium. For example, for the grand canonical ensemble the free energy is the $\Omega$-potential \cite{evans1979nature,Wu2006DensityMaterials, Roth2010FundamentalReview, Ravikovitch2001DensityNanopores}

\begin{equation}\label{eq:Omega}
    \Omega\left[\rho\left(\vec{r}\right)\right]=F\left[\rho\left(\vec{r}\right)\right]+\int d^3r \rho\left(\vec{r}\right)\left(V_{ext}\left(\vec{r}\right)-\mu\right),
\end{equation}
where $F\left[\rho\left(\vec{r}\right)\right]$ is the Helmholtz free energy, $\rho\left(\vec{r}\right)$ is a fluid density at the point $\vec{r}$, $V_{ext}$ is the external potential, $\mu$ is the chemical potential of fluid; temperature, volume and chemical potential are fixed.

The equilibrium density distribution $\rho\left(\vec{r}\right)$ meets the
following condition \cite{Wu2006DensityMaterials}
\begin{equation}\label{eq:variation}
    \dfrac{\delta\Omega\left[\rho\left(\vec{r}\right)\right]}{\delta \rho\left(\vec{r}\right)} = 
    \dfrac{\delta F\left[\rho\left(\vec{r}\right)\right]}{\delta \rho\left(\vec{r}\right)} + V_{ext}\left(\vec{r}\right) - \mu = 0. 
\end{equation}

In order to solve \eqref{eq:variation}, it is necessary to know the form of the Helmholtz energy and its variation. In this paper, we consider fluids nitrogen and argon; therefore, the Helmholtz energy can be divided into the sum of terms \cite{Wu2006DensityMaterials}:
\begin{equation}
    F\left[\rho\right] =  F^{id}\left[\rho\right]+ F^{ex}\left[\rho\right].\label{eq:sum_ener}
\end{equation}

The term $F^{id}$ in \eqref{eq:sum_ener} is the ideal part of free energy
\begin{equation}
    F^{id}\left[\rho\right] = k_B T \int d\vec{r}\,\rho\left(\vec{r}\right)\left(\ln{{(\Lambda}^3 \rho\left(\vec{r}\right))}-1\right),\label{eq:id_ener}
\end{equation}
where $k_B$ is the Boltzman's constant, $T$ is the temperature, $\Lambda = \frc{h}{\sqrt{2\pi mT}}$ is thethermal de Broglie wavelength, $h$ ---~Planck's constant, and $m$ is the gas molecule mass. 

The term $F^{ex}$ in \eqref{eq:sum_ener} is the excess part of free energy, which accounts for the intermolecular interactions.

In our work, we reproduce the excess part of free energy as the sum of hard-sphere and attraction interactions: 
\begin{align}
    F^{ex}\left[\rho\right] &= F^{hs}\left[\rho\right]+ F^{att}\left[\rho\right]\\
    F^{hs}\left[\rho\right] &= k_B T\int d\vec{r}\,\Phi^{RF}\left[n_\alpha\left(\rho\left(\vec{r}\right)\right)\right]\label{eq:HS_ener}\\
    F^{att}\left[\rho\right]&=k_B T\iint d\vec{r}\rho\left(\vec{r}\right)d\vec{r}^\prime\rho\left(\vec{r}^\prime\right)U_{att}(\vert\vec{r}-\vec{r}^\prime\vert) \label{eq:att_ener}.
\end{align}
The \eqref{eq:HS_ener} takes into account the hard sphere repulsion between fluid particles. The integrand is the Rosenfeld functional $\Phi^{RF}\left[n_\alpha\left(\rho\left(\vec{r}\right)\right)\right]$ with weighted density $n_\alpha\left(\rho\left(\vec{r}\right)\right)$ \cite{Rosenfeld1989Free-EnergyFreezing, Roth2010FundamentalReview}. Currently, there are several modifications to this functional. In this work, we use the classic form obtained by Rosenfeld according to Fundamental Measure Theory (FMT):
\begin{eqnarray}\label{eq:rosienfield}
    \Phi^{RF} =  -n_0 \ln{\left(1-n_3\right)} & + & \frac{n_1 n_2- \vec{n_1}\cdot\vec{n_2}}{1-n_3} \nonumber\\  & & +  \frac{n_2^3-3n_2 \vec{n_2}\cdot\vec{n_2}}{24\pi\left(1-n_3\right)^2},    
\end{eqnarray}
functions $n_\alpha, \bm{n}_\beta$ is the weighted densities ($\alpha = 1,2,3;\,\beta = 1,2$), expressions for them can be found in the appendix \ref{sec:Appendix_DFT}.

The \eqref{eq:att_ener} takes into account the contribution of the dipole-dipole interaction to the Helmholtz energy. Attraction potential $U_{att}$ is the Lennard-Jones potential, which is modified according to the WCA scheme \cite{Weeks1971RoleLiquids, Wu2006DensityMaterials}.

\begin{equation}
    U_{att}\left(r\right)=\ \left\{
    \begin{matrix}
        -\varepsilon_{ff}&r<\lambda\\
        U_{LJ}&\lambda<r<r_{cut}\\
        0&r>r_{cut}\\
    \end{matrix}\right.
\end{equation}
where $\lambda = 2^{1/6} \sigma_{ff}$ is the coordinate of the Lennard-Jones potential minimum, $\sigma_{ff}$ is the fluid particle diameter $\varepsilon_{ff}$ is the Lennard-Jones potential value at the minimum and $r_{cut}$ --- cut-off distance (in our work $r_{cut} = 5\sigma_{ff}$),
\begin{equation}
    U_{LJ}=4\varepsilon_{ff}\left(\left(\frac{\sigma_{ff}}{r}\right)^{12}- \left(\frac{\sigma_{ff}}{r}\right)^6\right).
\end{equation}

The external potential $V_{ext}$, which is included in the expression for the free energy \eqref{eq:Omega}, defines the interaction between the fluid molecules and the wall. In this paper we used the fluid-surface interaction potential of Steele 10-4-3 \cite{steele1974interaction}, the expression for it can be found in the appendix \ref{sec:Appendix_DFT}.
From the equality to zero of the $\Omega$-potential variation in equilibrium, taking into account \eqref{eq:sum_ener}--(\ref{eq:att_ener}), one can obtain an expression for the density
\begin{equation}\label{eq:density}
    \hat{\rho}\left(\vec{r}\right) = \rho^{bulk}\exp{\left\lbrace  c^{(1)}\left[\rho\left(\vec{r}\right)\right] - \beta V_{ext}\left(\vec{r}\right) + \beta\mu^{ex} \right\rbrace},
\end{equation}
where $\mu^{ex} = \mu^{att} + \mu^{hs}$ ---~chemical potentials, the expression of which can be found in the appendix~\ref{sec:Appendix_DFT}, $\beta = \frc{1}{k_B T}$ and $c^{(1)}$ ---~the one-body direct correlation function \cite{Ravikovitch2001DensityNanopores}:
\begin{equation}
    c^{\left(1\right)}\left[\rho\left(\bm{r}\right)\right] = -\dfrac{1}{k_B T}\dfrac{\delta F^{ex}\left[\rho\left(\bm{r}\right)\right]}{\delta \rho\left(\bm{r}\right)}
\end{equation}

It will be a different form of excess term $F^{ex}$ and its variation, according to the type of fluids molecule (the expression for the Helmholtz free energy variation are given in the appendix~\ref{sec:Appendix_DFT}). In the next section we will discuss our variation free approach, which may help to avoid direct calculation of correlation function.

In \eqref{eq:density} the density is implicitly included in the right-hand side through the Helmholtz energy variation. The Picard method is not the only method for solving nonlinear equation but it is the most popular and widely used for solving nonlinear equations, in particular, for DFT equations \cite{Roth2010FundamentalReview,mairhofer2017numerical}. The bulk fluid density is chosen as the initial approximation $\rho^{j=0} = \rho^{bulk}$ (the stability and the rate of convergence of the method depend on the choice of the initial guess). At the next iteration step, the solution is defined as the sum of solutions at the previous $\rho^j$ and current iterations $\hat{\rho}^j$ from \eqref{eq:density}:
\begin{equation}
    \rho^{j+1} = \left(1-\gamma\right)\rho^{j} + \gamma\hat{\rho}^j.
\end{equation}
The parameter $\gamma\in\left[0,1\right]$ determines the degree of confidence in the new decision on the next iteration, the higher it is, the faster the algorithm converges, but at too high values of $\gamma$, the solution will be unstable and will not converge. The Picard iteration method is highly reliable, however, it often works for a long time. Also, for this method, it is necessary to calculate the Helmholtz free energy variation, and it imposes some restrictions on the form of the Helmholtz energy. The complication of the model, such as taking into account the surface heterogeneity \cite{Aslyamov2017DensitySurfaces, Jagiello20132D-NLDFTCorrugation, Neimark2009QuenchedCarbons} or the Coulomb interaction \cite{Wu2006DensityMaterials}, leads to a change in the Helmholtz free energy form, which entails a change in the corresponding variations. It becomes necessary to change almost every term in \eqref{eq:density} and add new ones. In work \cite{Aslyamov2017DensitySurfaces}, it is clearly seen how taking into account the surface heterogeneity can strongly affect the expressions for variations, and what difficulties may arise in calculating the equilibrium density, even though the considered fluid is simple to model.

\section{Variation free method}\label{sec:VF-DFT}
Variation Free Density Functional Theory uses stochastic optimization methods to find the equilibrium fluid density instead of the classical approaches with Picard iteration or conjugate gradient descent. With the help of stochastic optimization methods, such as the genetic algorithm (GA) and particle swarm optimization (PSO), it is possible to find the equilibrium density $\rho^*\left(z\right)$ without direct calculation of one body direct correlation function and reduce the search time for a solution.
 
The form of the Helmholtz free energy is different for different systems, so as its variation. For some systems calculating the variation is a very laborious task due to the complex structure of the system's free energy. For example, for studying graphene nanobubbles by classical density functional and elasticity theories \cite{aslyamov2020model} it is important to know the energy of the whole system and density distribution at the same time because this system has additional constraints for fluid molecules. Another example, in work \cite{Aslyamov2017DensitySurfaces}, from the standard model of gas adsorption on an ideally smooth surface, they passed to the consideration of adsorption on a heterogeneous surface. This entailed significant changes in the variation of the Helmholtz energy. Also, the behavior of complex fluid with different types of interactions are interesting in some cases. For this purpose, the Statistical Associating Fluid Theory (SAFT) \cite{Papaioannou2016ApplicationIndustry,Lafitte2013AccurateSegments,xi2020efficient} is actively developed. Due to the consideration of chain, association and dispersion interactions in the molecule, additional terms appear in the free energy expression. All these interactions might be taken into account in a different way in Helmholtz free energy, and it is also necessary to calculate the variation for these additional terms. In addition to the fact that it is necessary to choose a physical model and the corresponding type of free energy, the problems under study are often endowed with specific geometric properties, and these properties dictate the type of coordinate system for DFT model \cite{Roth2010FundamentalReview, ludescher2021adsorption}. For some physical systems, it is not always possible to calculate the Helmholtz energy variation without any simplifications of the model (especially for systems with additional constraints), because of the free energy complex structure. The use of the VF-DFT will detour these barriers and get sufficiently accurate solutions.

\subsection{Basic patterns analysis}
\begin{figure*}[thb!]
    \centering

    \begin{subfigure}[thb!]{0.49\linewidth}
    \caption{Three different elements from $\bm{X}$ }
    \label{fig:basis_matrix}
    \includegraphics[scale=1]{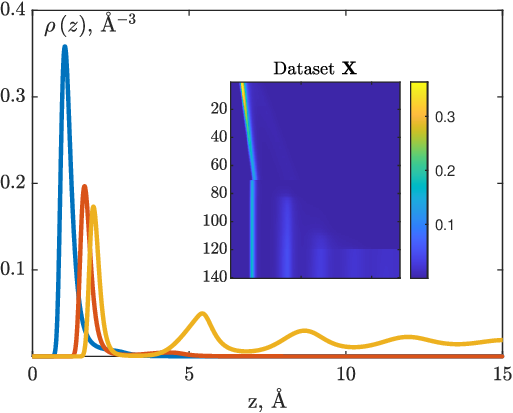}
    \end{subfigure}
\hfill
    \begin{subfigure}[thb!]{0.49\linewidth}
    \caption{Principlal components}
    \label{fig:PCA_basis}
    \includegraphics[scale=1]{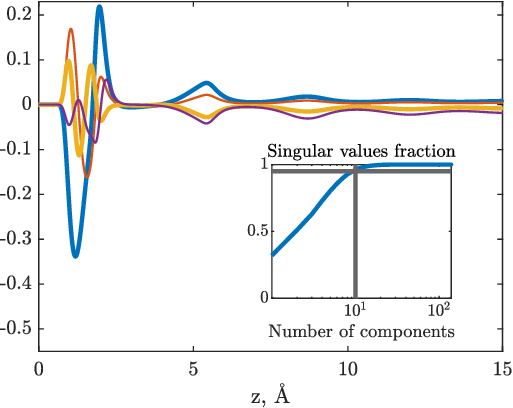}
    \end{subfigure}

    \caption{a) Three different elements $\rho\left(z\right)$ from dataset $\vec{X}$; dataset $\vec{X}$ which contains 140 different fluid density distributions. This dataset was used for building a set of basis functions. b) First four principle components from~$\vec{X}$ which were extracted by PCA for density decomposition. The relative error of dataset $\vec{X}$ reconstruction by first ten principle components $\sim 4\%$}
\end{figure*}
The information about the characteristic fluid density behavior in nanopore could be used to reduce the number of searched parameters, isolate the basic patterns, and seek the fluid density as a linear combination of functions: 
\begin{equation}
    \rho\left(z\right) = \sum\limits_{i = 1}^{K} a_i \phi_i \left(z\right)\label{eq:basis representation}
\end{equation}
In such a representation \eqref{eq:basis representation} of the sought function, the problem of finding the equilibrium density is reduced to finding the optimal expansion coefficients of $a_i$ that would minimize the $\Omega$-potential $\Omega\left[\rho\left(z\right)\right] = \Omega\left[a_i\right]$. The number and type of basic functions $\phi_i \left(z\right)$ will affect the quality of the solution and the speed of calculation 

For calculation basis functions $\phi_i \left(z\right)$ we used densities patterns, which was extracted from dataset $\bm{X}$ by PCA. A dataset $\bm{X}$ of nitrogen density functions at $ 77.4 $~K and different relative pressures (relation bulk pressure to saturation pressure from $ 10^{-6} $ to $0.99$) was created. These densities were computed by the classical DFT once for all considered cases in section~\ref{sec:results}. Also, in order to add as much information as possible about the characteristic fluid density behavior to the dataset $\bm{X}$, the equilibrium densities of "artificial" fluids with molecular radii from $0.85$~\AA,\ to $2$~\AA\ at the temperature $77.4$~K were taken. The other interaction parameters correspond to the parameters of nitrogen and have been fixed. As a result, we got a dataset $\vec{X}$ of 140 vectors as shown in \fref{fig:basis_matrix}. Such a selection of functions for the dataset does not limit the range of applicability of the algorithm. In section \ref{sec:results}, it will be demonstrated that the VF-DFT is applicable for fluid other than nitrogen with other thermodynamic conditions.

We applied the Principle Component Analysis (PCA) \cite{elizarev2020objective,mukhin2020application} to the dataset $\vec{X}$ for extracting the characteristic patterns of fluid density and for building basis functions. The advantage of this in reducing the number of optimized parameters without significant loss in quality of solution. After applying PCA to $\bm{X}$ we built 10 basis function which contains information about the main patterns of fluid behavior in nanopores.

Let the column vector $\vec{x}$ be one of the density distributions (there are 140 of them in our work). We normalized the matrix $\vec{X} (N_x \times N_s) $, $ N_x = 2401$ ---~dimension in $\rho$, $ N_s = 140$ ---~number of samples, as follows:

\begin{equation}
    \vec{X}\left(:,i\right) = \vec{x} - \vec{x}_{mean},
\end{equation}
where $\vec{x}_{mean} (N_x \times 1) $ is the average density distribution over the dataset $ \vec{X} $.

According to \cite{Abdi2010PrincipalStatistics}, we decompose the normalized $\vec{X}$ matrix using the Singular Value Decomposition (SVD):

\begin{equation}
    \vec{X} = \vec{U}\cdot\vec{\Sigma}\cdot \vec{V}^{\text{T}},
\end{equation}
where $ \vec{U}\left(N_x \times N_s \right) $ ---~the matrix of left singular vectors (orthonormal basis), $\vec{\Sigma}\left(N_s\times N_s\right) $ --~a square diagonal matrix containing singular numbers, sorted in descending order, $\vec{V}\left(N_s\times N_s \right)$ ---~a matrix of right singular vectors (also an orthonormal basis). As shown below, the left singular vectors contain information about patterns, and their corresponding singular values indicate how common this pattern in the dataset. To speed up the calculation in the PCA were applied truncated SVD, the number of components was selected by the energy criterion on $\vec{\Sigma}$. The square of the singular number is equal to the variance in the corresponding direction in the feature space. The projection in this direction will be minimal. The energy of the basis element is a square of the singular number, normalized to the sum of squares singular numbers
\begin{equation}
    I = \sigma^2/\sum\sigma^2.
\end{equation}
Respectively, the more energy, the better the selected components will describe the original dataset \cite{Sun2017AProblems,elizarev2020objective}.

Thus, $\vec{X}\approx \widetilde{\vec{X}}= \widetilde{\vec{U}}\cdot \widetilde{\vec{\Sigma}}\cdot\widetilde{\vec{V}}^{\text{T}}$. For the dataset $ \widetilde{\vec{X}} $ the sample covariance matrix can be calculated $\vec{Q}=\widetilde{\vec{X}}\cdot \widetilde{\vec{X}}^{\text{T}}/{(N_s-1)}$. Note that $\widetilde{\vec{X}}\cdot \widetilde{\vec{X}}^{\text{T}}=\widetilde{\vec{U}}\cdot\widetilde{\vec{\Sigma}}\cdot \widetilde{\vec{V}}^{\text{T}}\cdot \widetilde{\vec{V}}\cdot\widetilde{\vec{\Sigma}}^{\text{T}}\cdot \widetilde{\vec{U}}^{\text{T}}=\widetilde{\vec{U}}\cdot\widetilde{\vec{\Sigma}}^2 \cdot \widetilde{\vec{U}}^{\text{T}}$, i.e. eigenvalues of covariance matrix $\vec{Q}$ are equal to $\widetilde{\vec{\Sigma}}^2 /\left(N_s -1\right) $, and the eigenvectors of the covariance matrix $\vec{Q}$ are equal to the left singular vectors $\widetilde{\vec{U}}$. After calculation of the matrices $\widetilde{\vec{U}}$ and $\vec{S} = \vec{\widetilde{\vec{\Sigma}}}/ \sqrt{N_s - 1}$, the approach from \cite{Sarma2006EfficientUpdating} is used, where the new implementation is generated using the estimated covariance matrix as follows:
\begin{equation}\label{eq:new_dens}
    \vec{x}_{new} = \widetilde{\vec{U}}\cdot\vec{S}\cdot\vec{a} + \vec{x}_{mean}.
\end{equation}
Here $\vec{a}$ ---~vector $K\times 1 $ from the Gaussian distribution with the mean equal of zero and the standard deviation of one $\mathcal{N}\left(0,1 \right) $. Vector $\vec{a}$ can be interpreted as a vector of optimized parameters (components of this vector in fact are the expansion coefficients in \eqref{eq:basis representation} of the desired density). During the optimization, the components of the vector $\vec{a}$ and the vector $\vec{x}_{new}$ will be changing for minimizing the $\Omega$~potential.

As a result, after applying principal component analysis (PCA) on the matrix $\vec{X}$, a new basis of 10 vectors was calculated (the first five principal components is shown on \fref{fig:PCA_basis}), which cover 95\% of information by energy criterion. This means that instead of searching for 140 unknown numbers, we have to search for only 10 components of vector $\vec{a}$, and the quality of this approach will practically not differ from the solution with searching all 140 coefficients.

\subsection{Stochastic optimization}
As we have seen, the optimization problem is characterized by the fact that calculating the free energy variation is quite analytically tricky and change of the Helmholtz potential leads to the recalculation of each term in \eqref{eq:density}. In this work, was applied stochastic optimization methods. The optimized function was used as a black box, without calculating variations. The desired function was represented as \eqref{eq:new_dens}, and optimization algorithms sought the elements of the vector $\vec{a}$ to minimize the free energy functional $\Omega\left[\vec{a}\right]$. The dimension of the optimization problem is equal to the dimension of the vector $\vec{a}$. In our cases it is $10 \times 1 $. Were considered two iterative heuristic approaches---~GA and PSO, which are widely used in science and technology and have proven themselves in many industries \cite{NejadEbrahimi2013GeneticImages,Onwunalu2009DevelopmentDevelopment,Eberhart2001ParticleResources,Montes2001TheOptimization}. This section will provide a brief description of how the algorithms work and recommendations for configuring their parameters.

\subsubsection{Genetic Algorithm (GA)}

The creation of a genetic algorithm was inspired by Darwinian theory of evolution. The algorithm is based on the idea that the fittest individuals, which evolve, inherit traits from their parents, mutate, and it is they who survive ("survival of the fittest").
GA has been successfully applied to solve a wide variety of problems from various disciplines: feature selection in machine learning, image processing \cite{Huang2007AInformation,Ouellette2004GeneticDetection,Sharma2010GeneticExcitation}. It is not possible to describe all the modifications and methods developed on the basis of the classical genetic algorithm. The most complete presentation can be found in the articles \cite{NejadEbrahimi2013GeneticImages,Montes2001TheOptimization}.

\textit{Initialization}: we will call the vector $\vec{a}$ a chromosome (vector of optimized parameters), a set of chromosomes---~a generation. We recommend taking $N_c$ chromosomes equal to the number of parameters to be optimized, in our case, $ N_c = 10 $. First, all the chromosomes are set randomly within the given boundaries. So $\vec{a}$ is the vector of random variables from the $\mathcal{N}(0,\,1)$ in range [0;1], and the boundaries might be taken as~$\pm 4\sigma$. The objective function (fitness function) is calculated ---~in the minimization problem, the smaller it is, the more the chromosome is "adapted". In our case, the role of the objective function is played by the Omega potential $ \Omega\left[\vec{a}\right] $. In the genetic algorithm, each iteration is the process of forming a new generation from a current generation. This process takes place with the help of genetic operators. We used the most basic operators and will describe them in more detail.

\textit{Elitism}: one chromosome with the best objective function value is transferred to the new generation from the current generation without changes. This allows us at least not to lose the best solution, which ensures the loosely monotonous convergence of the algorithm.

\textit{Crossover}: each new chromosome ($N_c-1$) is obtained by crossing a pair of parents. The probability of selecting a current chromosome crossing generation can be either predetermined or depend on the fitness function (roulette wheel \cite{NejadEbrahimi2013GeneticImages}). The use of one or another approach changes the algorithm's exploration and exploitation properties. We used the index crossing approach. Several points in the current chromosomes are selected, and all genes with indices between the chosen points are exchanging occurs in blocks of genes.

\textit{Mutation}: a new chromosome obtained at the previous stage of crossing from a pair of parental ones undergoes mutation with a given probability, i.e. random changing one or more of its components. A high mutation rate slows down the convergence of the algorithm but makes it possible to find the global optimum more successfully. The level of mutation in this work was determined at the level of 0.5 ---~half of the population's chromosomes are subject to mutation.

After the new generation formation for all chromosomes, the fitness function is calculated. The algorithm stops when the allotted number of fitness calls is reached.

\subsubsection{Particle Swarm Optimization (PSO)}

Genetic algorithms today are one of the most popular tools for solving optimization problems. However, using this is not without its drawbacks. The solution converges to the optimum only after a large number of evaluations, while it is not guaranteed that the minimum will be the global. There is no clear criterion for stopping in case of insufficient mutation, the solution can be in a local minimum for a very long time, monotonic convergence is not guaranteed. All this leads to the emergence of new, more developed GA, the use of GA in conjunction with other algorithms, as well as the development of new stochastic approaches.

Particle Swarm Optimization is one such approach which was constructed on the base of GA. It was first developed in 1995 by Kennedy, Eberhart and Shi \cite{Kennedy1995ParticleOptimization} to study the behavior of a swarm of insects or schools of fish that move around space in search of food. Later it was simplified and applied to solve various optimization problems \cite{Banks2007ADevelopment}. Let us consider this method in more detail, using the terminology from \cite{Onwunalu2009DevelopmentDevelopment}.

\textit{Initialization}: the vector $\vec{a}$ is called a particle, the set of particles is a swarm (by analogy with a chromosome and a generation). As in GA, each particle stores its current position $\bm{x}_i^{k}$ (coordinates in search space) and the corresponding objective function value $F^{obj}\left(\bm{x}\right)$ (Omega potential in our case). In addition, as well as a velocity vector $\bm{v}_i^{k}$ (the same size as the position vector). In addition, for each particle, the best value of the objective functional is stored, at any iteration achieved by it and the corresponding position. The objective function ($\Omega$-potential) is the smaller, the better.

The number of particles in the work is set equal to the number of optimized coefficients (10 in our work), and the initial velocity is zero. The initial position of the particles is set from a uniform distribution.

PSO is an iterative algorithm in which the new position of particle $ i $ at iteration $ k + 1 $, $\vec{x}_i^{k+1}$, is determined by adding speed term $\vec{v}_i^{k+1}$ to $\vec{x}_i^{k}:\vec{x}_i^{k+1}=\mathbf{x}_i^{k}+\vec{v}_i^{k+1}$. The $ \vec{v}_i ^{k + 1}$ components are calculated as follows:

\begin{eqnarray}\label{eq:pso_velosity}
    \vec{v}_{i}^{k+1}=\omega \vec{v}_{i}^{k}&+&c_1\vec{r}_{1}\left({\hat{\vec{y}}}_{i}^{k} -\vec{x}_{i}^{k}\right)\nonumber \\ & & +c_2\vec{r}_{2}\left(\widetilde{\vec{y}}^{k}-\vec{x}_{i}^{k}\right),
\end{eqnarray}
where $\omega$, $c_1$, $c_2$ are some weights indicating the degree of attractiveness of a particular direction,  $\vec{r}$ is the random variables in the range from 0 to 1,  $\hat{\vec{y}}_{i}^{k}$ --- the best position for $i$ through passed iterations,  $\widetilde{\vec{y}}^{k}$ --- the best position through passed iterations among all particles.

It can be seen from the \eqref{eq:pso_velosity} that the speed is made up of three components, called inertial, cognitive and social, respectively. The inertial component $\omega \vec{v}_{i}^{k}$ is responsible for the movement of the particle in the direction in which it moved at the previous iteration $ k $. The cognitive term $c_1\vec{r}_{1}\left({\hat{\vec{y}}}_{i}^{k} -\vec{x}_{i}^{k}\right)$ is responsible for movement depending on the previous best position for the particle $ i $. The social component $c_2\vec{r}_{2}\left(\widetilde{\vec{y}}^{k}-\vec{x}_{i}^{k}\right)$ includes information about the best position for all particles and is responsible for moving towards it. At each iteration, each particle in the swarm is moved to a new location in the search space, and for each location an objective function is calculated that determines how good the given solution is.

The algorithm stops when it reaches the allocated number of calls of the objective function, and the best position of the particle is considered for all iterations response.

In this work, the coefficients $\omega=0.7298$, $c_1 = c_2=1.4962$. These values are the most versatile and suitable for many applications, it is worth noting that these three factors can not be arbitrary, and are related in \cite{Trelea2003TheSelection}. Changing these parameters did not give a better value of the objective function and did not reduce the time of calculation for our cases.

\subsection{Hybrid Density Functional Theory (H-DFT)}
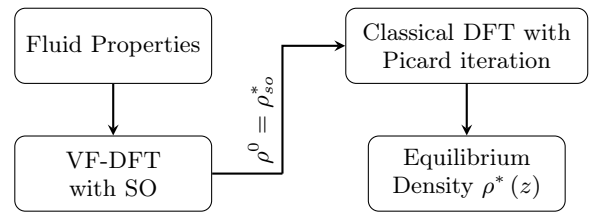
\begin{figure}[htbp]
\centering
\begin{tikzpicture}[node distance=1.7cm]
\node (start) [cdft,text width=2.4cm] {Fluid Properties};
\node (VF-DFT) [cdft,below of=start,text width=2.4cm] {VF-DFT with SO};
\node (DFT) [cdft,right of=start,xshift=3cm,text width=3cm] {Classical DFT with Picard iteration};
\node (dens) [cdft,below of=DFT,text width=2.4cm] {Equilibrium Density $\rho^{*}\left(z\right)$};

\draw [arrow] (start) -- (VF-DFT);
\draw [thick] (VF-DFT) -- +(2.25,0);
\draw [arrow] +(2.25,-1.7) node[anchor=south west,rotate=90] {$\rho^0 = \rho^*_{so}$} |- (DFT);
\draw [arrow] (DFT) -- (dens);

\end{tikzpicture}
\caption{Scheme of Hybrid Density Functional Theory}
\label{fig:H-DFT}
\end{figure}

The VF-DFT approach does not require the calculation of the Helmholtz free energy variation and works faster than the classical DFT with Picard iteration. However, classical DFT provides a more accurate solution. To maintain the advantages in computation speed up and quality of the solution from both approaches, we combined them into a single approach called Hybrid Density Functional Theory (H-DFT) \fref{fig:H-DFT}. At the initial stage of H-DFT, the approximate solution is sought by VF-DFT  with stochastic optimization techniques. Then, the solution produced by VF-DFT is supplied as the initial approximation for the classical DFT Picard iteration to clarify the solution. This combination made it possible to obtain a significant gain in the speed of density calculation, in comparison with the classical DFT, while the quality of the solution remained at the same level as in the classical algorithm. Due to the fact that the VF-DFT solution is close to equilibrium, the value of $\gamma$ for Picard iteration could be set large enough. This affects the speed of the Picard iteration and the number of iterations required for the algorithm. 

\section{Results}\label{sec:results}
This section presents the results for VF-DFT and H-DFT methods. GA and PSO are used for seeking optimal coefficients of expansion for desired fluid density in the pore. The developed methods were compared with the classical DFT in terms of calculation time and the value of $\Omega$-potential. Methods are tested for nitrogen and argon in the pore of 3.6 nm. Fluid-fluid and solid-fluid interaction parameters represented in the table \ref{tab:fluid-param}. The parameters for solid-fluid interaction were calculated according to the Lorentz-Berthelot rule \eqref{eq:Lor-Bert}, where $\epsilon_{ss} = 28$~K, $\sigma_{ss} = 3.4$~\AA
\begin{equation}\label{eq:Lor-Bert}
    \epsilon_{sf} = \sqrt{\epsilon_{ss}\epsilon_{ff}}, \quad \sigma_{sf} = \dfrac{1}{2}\left(\sigma_{ss} + \sigma_{ff}\right).
\end{equation}

\renewcommand{\arraystretch}{1.1} 
\renewcommand{\tabcolsep}{0.2cm} 
\begin{table}[htb!]
\caption{The fluid-fluid and solid-fluid interaction parameters. For LJ potential cut off distance was taken $r_{cut} = 5\sigma_{ff}$. Surface density $\rho_V = 0.114$ \AA$^{-3}$, distance between layers of carbon atoms $\Delta = 3.35$ \AA}
\label{tab:fluid-param}
\begin{ruledtabular}
\begin{tabular}{@{}clcccc@{}}

\multicolumn{2}{c}{}                                 & \multicolumn{2}{c}{Fluid-Fluid} & \multicolumn{2}{c}{Solid-Fluid} \\ \midrule
\multirow{2}{*}{Fluid} &
  \multirow{2}{*}{$T,$ K} &
  \multirow{2}{*}{$\frc{\epsilon_{ff}}{k_B},$ K} &
  \multirow{2}{*}{$\sigma_{ff},$ \AA} &
  \multirow{2}{*}{$\frc{\epsilon_{sf}}{k_B}$, K} &
  \multirow{2}{*}{$\sigma_{sf},$ \AA} \\
                          &                          &                 &               &                &                \\
\multicolumn{1}{l}{$N_2$} & \multicolumn{1}{c}{77.4} & 94.45           & 3.575         & 51.43          & 3.487          \\
\multicolumn{1}{l}{$Ar$}  & \multicolumn{1}{c}{87.3} & 111.95          & 3.358         & 55.99          & 3.379          \\ 
\end{tabular}
\end{ruledtabular}
\end{table}


\subsection{Basis functions and problem statement for identical fluids}

Nitrogen was taken in the first stage of testing the algorithm because the dataset contains the most information about this fluid. As already mentioned, the basis functions were constructed from the dataset of equilibrium nitrogen densities $\bm{X}$ in the pore of $3.6$~nm at the fixed temperature of $77.4$~K and different relative pressures. The dataset $\bm{X}$ also include the equilibrium densities of “artificial” fluids with different molecular radii (from $0.85$~\AA\ to $2$~\AA) with the rest of the interaction parameters fixed. For all our calculations we used the same basis functions.

\begin{figure*}[htb!]
    \centering

    \begin{subfigure}[htb!]{0.49\linewidth}
    \caption{VF-DFT: density for high pressure case with Nitrogen}
    \label{fig:Density_hp_N}
    \includegraphics[scale=1]{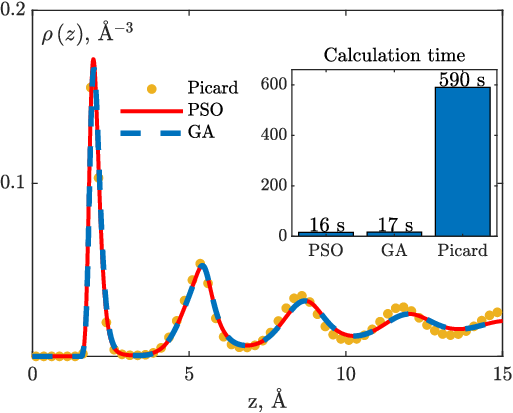}
    \end{subfigure}
\hfill
    \begin{subfigure}[htb!]{0.49\linewidth}
    \caption{H-DFT: density for high pressure case with Nitrogen}
    \label{fig:Density_hp_N_C}
    \includegraphics[scale=1]{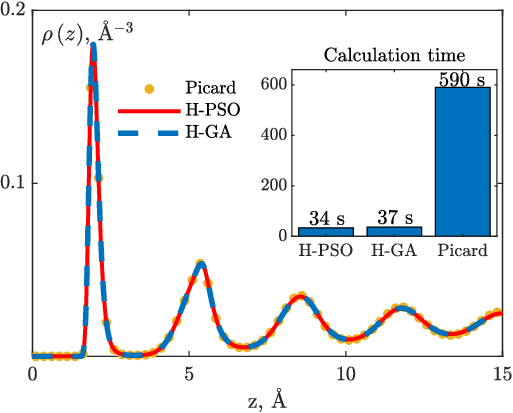}
    \end{subfigure}
\hfill
    \begin{subfigure}[htb!]{0.49\linewidth}
    \caption{VF-DFT: density for low pressure case with Nitrogen}
    \label{fig:Density_lp_N}
    \includegraphics[scale=1]{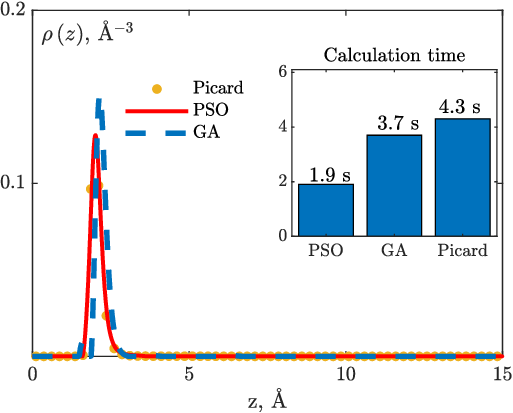}
    \end{subfigure}
\hfill
    \begin{subfigure}[htb!]{0.49\linewidth}
    \caption{H-DFT: density for low pressure case with Nitrogen}
    \label{fig:Density_lp_N_C}
    \includegraphics[scale=1]{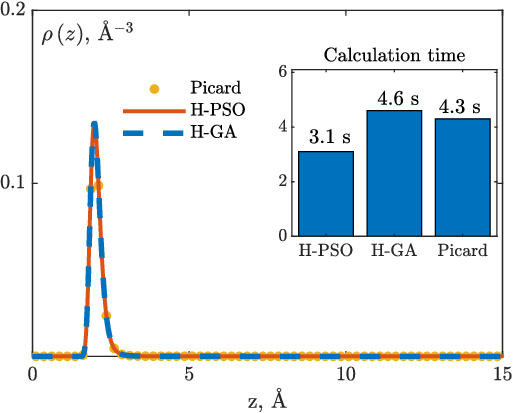}
    \end{subfigure}

    \caption{Red solid line ---~VF-DFT (H-DFT) with PSO, blue meshed line ---~VF-DFT (H-DFT) with GA, yellow circles indicate classical DFT with Picard iteration. a) VF-DFT equilibrium densities at $P/P_0 = 0.6924$ for nitrogen $\Omega_{PSO} = -0.2630$,  $\Omega_{GA} = -0.2628$, $\Omega_{Pic} = -0.2634$. b) H-DFT equilibrium densities at $P/P_0 = 0.6924$ for nitrogen. The figures completely coincide, but the running time of hybrid methods is an order of magnitude less $\Omega_{H-PSO} =\Omega_{H-GA} =\Omega_{Pic} =-0.2634$. c) VF-DFT equilibrium densities at $P/P_0 = 0.0044$ for nitrogen $\Omega_{PSO} = -0.0305$, $\Omega_{GA} = -0.0286$, $\Omega_{Pic} = -0.0308$. d) H-DFT equilibrium densities at $P/P_0 = 0.0044$ for nitrogen $\Omega_{H-PSO} =\Omega_{H-GA} =\Omega_{Pic} =-0.0308$
    }
\end{figure*}

\paragraph{High pressure.}
Classical algorithms are looking for an equilibrium solution for a long time, when calculating the equilibrium density at large values of relative pressure. It is in such problems that the variation free approach gives the most significant gain in computation time. Even though Picard iteration is very reliable, to ensure convergence in such problems, the $\gamma$ parameter should be chosen very small, which dramatically affecting the convergence rate. The \fref{fig:Density_hp_N} shows the nitrogen density in a pore of 3.6 nm at a temperature of 77.4 K and a relative pressure in the bulk $ P/P_0 = 0.6924 $ ($P_0$ is the saturation pressure). It is important to note that the dataset does not include densities at the relative pressures at which calculations are made in the cases under consideration. The difference in calculation time is significant, however, the solution is slightly different in the value of the $\Omega$~potential to the classical DFT.

VF-DFT algorithms with GA, PSO managed with the task an order of magnitude faster, but the solution is slightly different from the classical approach. However, the structural feature of fluid behavior was reproduced well by the VF-DFT, the number of peaks and their shapes are in good agreement with the solution of the classical approach. PSO-based VF-DFT provided the value of the $\Omega$~potential for the calculated equilibrium density $\Omega\left[\rho^*_{PSO}\right] = -0.2630$, GA-based VF-DFT $\Omega\left[\rho^*_{GA}\right] = -0.2628$, both turned out to be very close to the value given by the DFT with the Picard iteration method $\Omega\left[\rho^*_{Pic}\right] = -0.2634$. In section III A was mentioned that in the VF-DFT approach the density is represented as expansion through the limited set of basis functions $\phi_i\left(z\right)$. That is why in all our cases variation free approach gave the solution different from classical DFT with Picard iteration. To reduce the difference between the classical approach and VF-DFT might be used different basis functions or our Hybrid Density Functional Theory method. 

The \fref{fig:Density_hp_N_C} shows the result of hybrid methods versus Picard iteration. The running time is slightly different from the VF-DFT \fref{fig:Density_hp_N}, but the final value of the $\Omega$~potential for all three density profiles is the same $\Omega\left[\rho^*_{H-PSO}\right] =\Omega\left[\rho^*_{H-GA}\right] =\Omega\left[\rho^*_{Picard}\right] =-0.2634$.

\paragraph{Low pressure.}
The \fref{fig:Density_lp_N} shows that at low relative pressures $P/P_0 = 0.0044$, the computation speed for VF-DFT is not as significant as at high pressures. At low relative pressures, classical algorithms work faster than at high pressures, due to the choice of the parameter $\gamma$ large enough.

For hybrid algorithms \fref{fig:Density_lp_N_C}, it was turned out that the genetic algorithm is slower than Picard iteration, VF-DFT with PSO is still faster than both, and the value of the $\Omega$ ~ potential for all three is the same $\Omega\left[\rho^*_{H-PSO}\right] =\Omega\left[\rho^*_{H-GA}\right] =\Omega\left[\rho^*_{Pic}\right] =-0.0308$.

In this case, the speed of the algorithms is significantly affected by the initialization time of the optimization algorithms, and for hybrid algorithms this time is longer, since, in addition to stochastic PSO and GA, it is required to initialize Picard iteration for which the free energy variation must also be calculated.

\begin{figure}[h]
    \centering
    \includegraphics{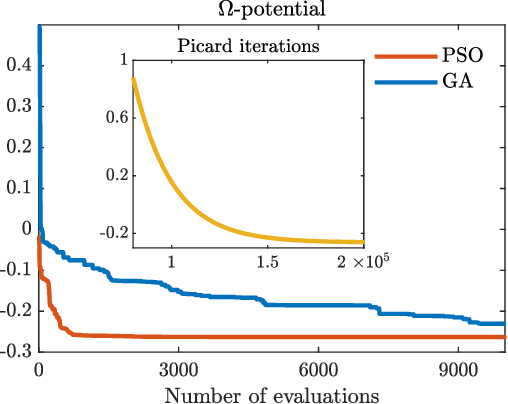}
    \caption{
    The value of $\Omega$~potential with iterations for classical DFT (yellow solid line), and $\Omega$~potential with evaluation for VF-DFT with PSO (red solid line) and GA (blue dashed line)
    }
    \label{fig:Convergence_hp_N}
\end{figure}

The value of the $\Omega$-potential for the Picard iteration method decreases much more slowly compared to the VF-DFT, as seen in \fref{fig:Convergence_hp_N}. 
 
The behavior of the $\Omega$~potential for stochastic algorithms \fref{fig:Convergence_hp_N} is not strictly monotone. This is due to the principle of the algorithms themselves. There are no exact convergence criteria for stochastic algorithms, which means that the solution that the algorithm produces does not have to be acceptable. In this case, the calculation have to be continued with hope that the solution will be improved or run the algorithm again. Stability and speed of search procedure depends on settings of algorithm. Also, it should be noted the difference between iterations and evaluations for algorithms. For example, if we have a population of $Nc$ chromosomes, each of them gets evaluated once in a single iteration. This means that in every iteration of an algorithm, the evaluation function is called $Nc$ times (once for each chromosome). Hence, we have the relationship: $\textit{Number of evaluations} = \textit{Number of iterations} \cdot Nc
$.

Variation free approach makes it possible to obtain a solution significantly faster. The combination of VF-DFT with classical DFT provides (H-DFT), in addition to high-speed computing the same quality solutions as the classical approach.
\begin{figure*}[htb!]
    \centering
\begin{subfigure}[htb!]{0.49\linewidth}
    \caption{VF-DFT: density for high pressure case with Argon}
    \label{fig:Density_hp_A}
    \includegraphics[scale=1]{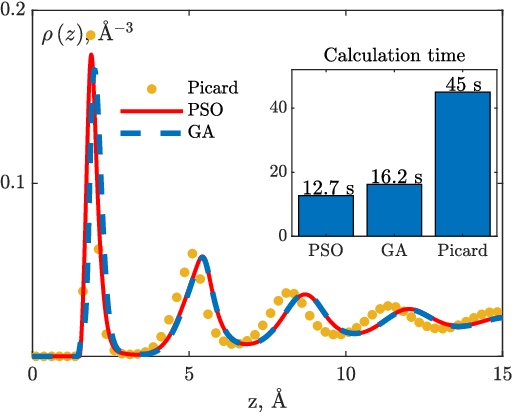}
    \end{subfigure}
\hfill
    \begin{subfigure}[htb!]{0.49\linewidth}
    \caption{H-DFT: density for high pressure case with Argon}
    \label{fig:Density_hp_A_C}
    \includegraphics[scale=1]{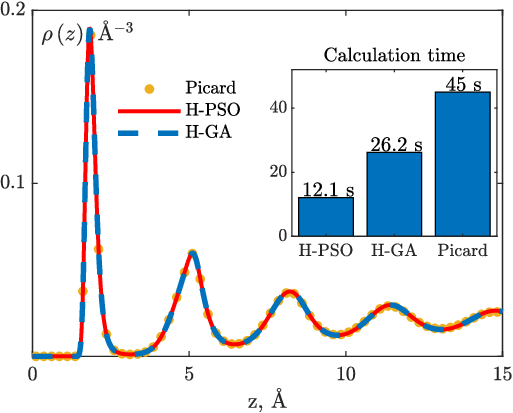}
    \end{subfigure}
\hfill
    \begin{subfigure}[htb!]{0.49\linewidth}
    \caption{VF-DFT: density for middle pressure case with Argon}
    \label{fig:Density_mp_A}
    \includegraphics[scale=1]{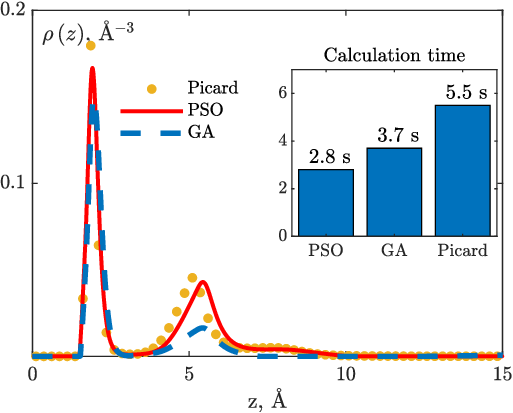}
    \end{subfigure}
\hfill
    \begin{subfigure}[htb!]{0.49\linewidth}
    \caption{H-DFT: density for middle pressure case with Argon}
    \label{fig:Density_mp_A_C}
    \includegraphics[scale=1]{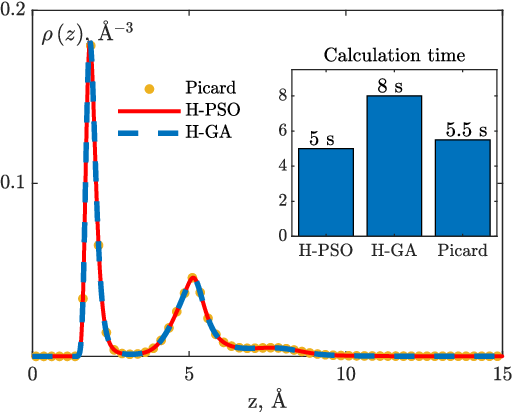}
    \end{subfigure}

    \caption{Red solid line ---~VF-DFT (H-DFT) with PSO, blue meshed line ---~VF-DFT (H-DFT) with GA , yellow circles indicate classical DFT with Picard iteration a) VF-DFT equilibrium densities at $P/P_0 = 0.9932$ for argon $\Omega_{PSO} = -0.1674$, $\Omega_{GA} = -0.1626$, $\Omega_{Pic} = -0.1712$ b) H-DFT equilibrium densities at $P/P_0 = 0.9932$ for argon $\Omega_{H-PSO} =\Omega_{H-GA} =\Omega_{Pic} =-0.1077$. c) VF-DFT equilibrium densities at $P/P_0 = 0.4739$ for argon $\Omega_{PSO} = -0.1049$, $\Omega_{GA} = -0.1019$, $\Omega_{Pic} = -0.1077$ d) H-DFT equilibrium densities at $P/P_0 = 0.4739$ for argon $\Omega_{H-PSO} =\Omega_{H-GA} =\Omega_{Pic} =-0.1077$
    }\label{fig:diff_all}
\end{figure*}

\subsection{Basis functions and problem statement for different fluids}

This section represents the most valuable results from the VF-DFT \fref{fig:diff_all}. To make sure that VF-DFT can be used not only for calculating the equilibrium nitrogen density at the temperature of $77.4$~K (the basis was considered based on nitrogen), but also for other fluids, we applied VF-DFT for argon at the temperature of $87.3$~K. The results of this section confirm the idea of VF-DFT applicability for calculating the density of a fluid with a complex structure of Helmholtz free energy without calculating variations.

\paragraph{High pressure.}

The \fref{fig:Density_hp_A} shows that the position of the peaks did not coincide exactly, but it is worth noting that the number and size of the peaks VF-DFT was able to restore. Also, VF-DFT still works many times faster than the classic approach. As was mentioned in cases with nitrogen the difference in solutions between VF-DFT and classical approach by reason of the limited basis functions set. The difference between solutions might be reduced by adding other functions to the basis or applying the hybrid method.

When using the hybrid algorithm \fref{fig:Density_hp_A_C}, the solutions completely coincided with a shorter calculation time.

\paragraph{Middle pressure.}

As the relative pressure decreases, the VF-DFT provide a smaller gain in speed \fref{fig:Density_mp_A}. In this case, the stochastic algorithms were able to catch the position of the peaks, but only VF-DFT with PSO was able to recover the magnitude of the peaks.

The hybrid approach in the case with relative pressure $P/P_0 = 0.4739$ \fref{fig:Density_mp_A_C} did not give the expected gain in the calculation speed. The GA-based H-DFT turned out to be even slower than the classical DFT. As already was mentioned in the section with the nitrogen results, this is due to the initialization time for optimization algorithms. In this case, a sufficiently large value of $\gamma$ can be used for the Picard iteration. The time spent in the H-DFT to obtain an approximate solution turned out to be very close to the time of the classical approach.

In all considered cases, the greatest gain in speed was obtained at high relative pressures. This is due to the fact that it is possible to choose a sufficiently small parameter $\gamma$ for the convergence of the Picard iteration method at high relative pressures. In addition, at high relative pressures, more layers are formed near the wall, which means more patterns can be extracted by PCA. Thus, it is easier to look for a solution using VF-DFT at high pressures.

\section{Conclusion}
Density functional theory is a powerful tool for investigating the behavior of fluids and surfaces, taking into account molecular forces at the nanoscale. The Variation Free approach developed in this work for seeking the equilibrium density is of interest for the study of intricate systems: systems with additional constraints, fluids with complex interactions. Besides, the combination of VF-DFT, which uses stochastic optimization methods, with the classical DFT with Picard iteration (H-DFT approach) made it possible to significantly speed up the calculation for high relative pressure cases while maintaining the solution quality at the same level as in the classical approach. At low relative pressures, the calculation speed gain is not as significant as at high pressures because of specific pattern extraction from the dataset. The combined algorithm can be applied to speed up calculations of the equilibrium fluid density at high pressures, in particular, to speed up the calculation of the adsorption isotherm or pore stresses. It should be noted that the VF-DFT solution's speed and quality directly depend on the basis functions. The number of the basis functions and their type can be changed, improved, and refined to obtain a better solution with a minimum value of calculation time. In addition to the basis, the speed of VF-DFT operation is influenced by the settings of the optimization algorithms. The optimization algorithms GA and PSO considered in this work showed promising results, PSO has a higher quality and speed of finding the solution. In the future, it is possible to investigate the variation free approach with other optimization methods, investigate systems for which the Helmholtz free energy has a complex form or systems with specific constraints.

\begin{acknowledgments}
The authors are grateful to Timur Aslyamov for critical comments and fruitful discussions about this work.
\end{acknowledgments}


\appendix

\section{Details for Density Functional Theory}

\label{sec:Appendix_DFT}
The \eqref{eq:rosienfield} on the page \pageref{eq:rosienfield} contains weighted density functions $n_\alpha$, which are defined as follows:
\begin{equation}\label{eq:weighted_dens}
    n_\alpha \left(\vec{r}\right)=\int d^3r^\prime \rho\left(\vec{r}^\prime\right)\omega_\alpha\left(\vec{r}-\vec{r}^\prime\right),
\end{equation}
where $\omega_\alpha$ --- weight functions; $\omega_3\left(\vec{r}\right)=\theta\left(R-r\right)$, $\omega_2\left(\vec{r}\right)=\delta\left(R-r\right)$, ${\vec{\omega}}_2\left(\vec{r}\right)=\frac{\vec{r}}{r}\delta\left(R-r\right)$, $\omega_1= \frac{\omega_2}{4\pi R}$, $\omega_0 = \frac{\omega_2}{4\pi R^2}$, ${\vec{\omega}}_1= \frac{{\vec{\omega}}_2}{4\pi R}$, $\delta$ и $\theta$ --- Dirac delta function and Heaviside theta function, $R$ --- fluid particle radius.\\

Similarly, the Helmholtz energy \eqref{eq:sum_ener}, the chemical potential of the fluid can also be divided into the sum of three components

\begin{align}
    \mu\left(\rho\right)&= \mu_{id}\left(\rho\right)+\mu_{HS}\left(\rho\right)+\mu_{att}\left(\rho\right)\\
    \mu_{id}\left(\rho\right)&=k_B T \left(\ln{\lambda^3 \rho^{bulk}}\right)=const\\
    \mu_{HS}\left(\rho\right) &=k_B T\left(\sum{\frac{\partial\Phi}{\partial n_\alpha}\frac{\partial n_\alpha }{\partial\rho}}\right)=const\\
    \mu_{att}\left(\rho\right)&=\rho^{bulk}\int{d^3r\, U_{att}\left(r\right)}=const \label{eq:chem_att}
\end{align}
where $\rho^{bulk}$---~fluid density in the bulk.\\

In this paper, the pores of the slit geometry are considered. The problem is symmetric in two directions ($ x,\ y $), and the density depends only on one coordinate $z$: $\rho\left(\vec{r}\right) = \rho\left(z\right)$. The pore walls are composed of carbon atoms. The potential of fluid-solid interaction is described by the Still potential 10-4-3 \cite{steele1974interaction}.
\begin{equation}
\begin{split}
V_{sf}\left(z\right) =  2\pi\varepsilon_{sf}\rho_V\sigma_{sf}^3\Delta & \left(\frac{2}{5}\left(\dfrac{\sigma_{sf}}{z}\right)^{10} - \left(\dfrac{\sigma_{sf}}{z}\right)^4 - \right.\\
    & \left. - \dfrac{\sigma_{sf}^4}{3\Delta\left(0.61\Delta+z\right)^3}\right)
\end{split}
\end{equation}
Full potential in the pore from two walls:
\begin{equation}
    V_{ext}\left(z\right)= V_{sf}\left(z\right)+ V_{sf}\left({H}_{cc}-z\right), 
\end{equation}
$\rho_V$ --- the density of the wall material, $\Delta$ --- the distance between the layers of wall atoms, $H_{cc}$ --- pore diameter calculated between the centers of the carbon atoms of opposing sides, the width of the pores accessible to fluid $H=H_{cc}-2z_0+\sigma_{ff}$, where $z_0$ --- penetration rate ( $z_0 = 0.9\sigma_{sf}$).

In the case of a planar geometry could be write down analytic form Helmholtz free energy variation, for hard sphere repulsion:
\begin{equation}
    \dfrac{\delta F^{HS}\left[\rho\right]}{\delta\rho\left(z\right)} = \sum\limits_\alpha \int dz' \dfrac{\partial\Phi\left[n_\alpha\right]}{\partial n_\alpha}\dfrac{\delta n_\alpha\left(z'\right)}{\delta \rho\left(z\right)}
\end{equation}
\begin{equation}
    \dfrac{\delta n_\alpha\left(z'\right)}{\delta \rho\left(z\right)} = \dfrac{\delta}{\delta \rho\left(z\right)}\int dz'' \rho\left(z''\right)\omega_{\alpha}\left(z'-z''\right) = \omega_{\alpha}\left(z' - z\right)
\end{equation}
Free energy variation for attraction interaction:
\begin{equation}
    F^{att}= \frac{k_B T}{2}\iint dz_1 dz_2\,\rho\left(z_1\right)\rho\left(z_2\right)G\left(z_1,z_2\right), 
\end{equation}
\begin{widetext}
where, taking into account the type of potential attraction interaction $U_{att}\left(r\right)$ при $\vert \Delta z\vert \le \lambda$
\begin{multline}
    G\left(z_1,z_2\right) = 2\pi\left\{\int\limits_{0}^{\sqrt{\lambda^2-\Delta z^2}}{-\varepsilon_{ff}rdr+\ \int\limits_{\sqrt{\lambda^2-\Delta z^2}}^{\sqrt{r_{cut}^2-\Delta z^2}}{dr\ 4\varepsilon_{ff}r\left[\frac{\sigma_{ff}^{12}}{\left(\Delta z^2+r^2\right)^6}-\ \frac{\sigma_{ff}^6}{\left(\Delta z^2+r^2\right)^3}\right]}}\right\} =\\=
    -\pi\varepsilon_{ff}\left(\lambda^2-\Delta z^2\right)+\frac{4}{5}\pi\varepsilon_{ff}\sigma_{ff}^{12}\left(\frac{1}{\lambda^{10}}-\ \frac{1}{r_{cut}^{10}}\right)-2\pi\varepsilon_{ff}\sigma_{ff}^6\left(\frac{1}{\lambda^4}-\ \frac{1}{r_{cut}^4}\right),
\end{multline}
where $\vert \Delta z\vert > \lambda$ and $\vert \Delta z\vert \le r_{cut}$
\begin{multline}
    G\left(z_1,z_2\right) = 2\pi\int\limits_{0}^{\sqrt{r_{cut}^2-\Delta z^2}}{dr\ 4\varepsilon_{ff}r\left[\frac{\sigma_{ff}^{12}}{\left(\Delta z^2+r^2\right)^6}-\ \frac{\sigma_{ff}^6}{\left(\Delta z^2+r^2\right)^3}\right]} =\\=
    \frac{4}{5}\pi\varepsilon_{ff}\sigma_{ff}^{12}\left(\frac{1}{{\Delta z}^{10}}-\ \frac{1}{r_{cut}^{10}}\right)-2\pi\varepsilon_{ff}\sigma_{ff}^6\left(\frac{1}{{\Delta z}^4}-\ \frac{1}{r_{cut}^4}\right),
\end{multline}
with $\vert \Delta z\vert > r_{cut}$, $G\left(z_1,z_2\right) = 0$.
\end{widetext}
Thus, the variation of the Helmholtz energy from attraction interaction
\begin{equation}
    \dfrac{\delta F^{att}\left[\rho\left(z\right)\right]}{\delta \rho\left(z\right)} = \dfrac{k_B T}{2} \int dz'\, \rho\left(z'\right)G\left(z - z'\right)
\end{equation}

\selectlanguage{english}
\bibliography{references}

\end{document}